# Modularity of Biochemical Filtering for Inducing Sigmoid Response in Both Inputs in an Enzymatic AND Gate

Saira Bakshi,[a] Oleksandr Zavalov,[b] Jan Halámek,[a] Vladimir Privman,[b] Evgeny Katz[a]

[a]*Department of Chemistry and Biomolecular Science, and*
[b]*Department of Physics, Clarkson University, Potsdam, NY 13676*

**Abstract.** We report the first systematic study of designed two-input biochemical systems as information processing gates with favorable noise-transmission properties accomplished by modifying the gate's response from convex shape to sigmoid in both inputs. This is realized by an added chemical "filter" process which recycles some of the output back into one of the inputs. We study a system involving the biocatalytic function of the enzyme horseradish peroxidase, functioning as an **AND** gate. We consider modularity properties, such as the use of three different input chromogens that, when oxidized yield signal-detection outputs for various ranges of the primary input, hydrogen peroxide. We also examine possible uses of different filter-effect chemicals (reducing agents) to induce the sigmoid-response. A modeling approach is developed and applied to our data, allowing us to describe the enzymatic kinetics in the framework of a formulation suitable for evaluating the noise-handling properties of the studied systems as logic gates for information processing steps.

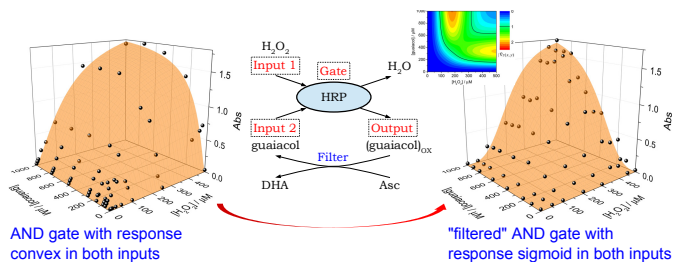

## 1. Introduction

There has been a substantial body of work in the chemical[1-6] and biochemical literature[7-12] on interpreting various simple (bio)chemical reactions in terms of binary logic gates. This research has been motivated by the promise of applications in devising new computation,[13-15] sensing,[16-18] and more generally information transmission and processing approaches,[19-21] as well as adding new functionalities[22-24] to conventional electronics. Biochemical information processing[7-12] has been of particular interest as a promising approach to expand the biocompatible interfacing capabilities of electronic devices,[25-30] and also in utilizing the selectivity and richness of the biomolecules for handling information and signals in complex settings.[31-33] Basic binary gates, such as **AND**, **OR**, **XOR**, etc. have been realized,[7-12,34] some aiming at specific diagnostic applications,[35-41] and small networks were demonstrated.[19,20,31,32] This was done with proteins/enzymes,[12,42] DNA,[19,20,43,44] RNA[45] and whole cells.[46-48] Our focus here is on enzyme-catalyzed processes because of the their wide utilization in biosensing,[16-18] medical diagnostics[35-41] and interfacing with electronic devices.[25-30]

Future applications will aim at achieving versatile and complex information processing with biomolecules, necessitating networking of biocomputing gates, which could lead to noise amplification. For large-scale networking and fault-tolerant information processing within the binary electronics paradigm, the simple two-input, single-output gates, here exemplified by **AND**, have to be properly designed[49] (optimized) to avoid too much noise amplification and preferably achieve analog noise suppression[50] in most of the networked steps. This requires consideration of the gate's "response function:" the output as a function of its inputs not only at the values selected as the reference binary "logic" **0** and **1**, but also in the ranges between these values and when physically relevant, somewhat beyond these ranges, as explained later (in Section 3). A successful recent approach has involved "biochemical filtering," i.e., simple chemical modifications of the biocatalytic enzymatic processes to make the response function sigmoid, preferably in both inputs.[51-60] Other approaches, specifically, those utilizing allosteric enzymes with self-promoter substrates have also been considered.[61-64]

Approaches to "biochemical filtering" include[51-60] the use of an added chemical process to consume/deactivate a fraction of the input[57-60] or output[54] in order to decrease the magnitude of the observed output signal at small input(s) and output, i.e., near logic-**0**(s). The added process involves a reactant that is exhausted after consuming some of the targeted signal-chemicals, and therefore at larger signals, at logic-**1**(s), the effect of the introduced "filtering" is limited. However, such filtering can reduce the overall output intensity range. One way to partially avoid this loss of intensity has been to use processes that instead of consuming a part of the output, recycle it into some intermediate chemical or even an input.[51,53,55,56]

The latter approach, with the added filtering process recycling the output rather than affecting one of the inputs, promises two advantages. The more obvious one is that the loss of the overall signal intensity during the gate function is partially avoided (by recycling). The other possible advantage is that the response can possibly be made sigmoid in two inputs by using a single added filtering reaction. The first expectation has been confirmed by the few presently reported experiments.[51,53,56] However, the latter possibility, that "filtering" the output of the two-input, single-output gate (here, the **AND** logic gate) can under proper conditions and with proper optimization by parameter selection yield a double-sigmoid (sigmoid in both inputs) response has not yet been studied beyond some preliminary evidence for **AND** and **OR** binary gates.[51,53,56]

In this work we report the first systematic modeling based on new experimental realizations of the "output filtering" involving recycling a part of the output into the second input of enzymatic **AND** gates based on the biocatalytic functioning of horseradish peroxidase (HRP), as defined in Section 3, with Sections 2 and 4 offering experimental details. This allows us to explore the degree of versatility of this approach by studying systems with different input chemicals and also with varying filtering-process compounds. We consider to what extent this filtering approach can be used to obtain a double-sigmoid response in situations when modularity is expected. By "modularity" we mean the ability to use the gate as a sub-process in various settings involving its networking with other (bio)chemical reactions. While some of the chemical parameters internal to the gate's functioning can be adjusted as needed, we have to be able to use different inputs and output according to the desired chemical steps in the network. Specifically, we use three different chemicals for the second (recycled) input of the **AND** gate, and for the optimal choice of this input, we also consider three different reactants that cause the filtering effect. These results, illustrated for our optimal case in Figure 1, are detailed in Section 4, whereas Section 5 offers the concluding comments.

## 2. Experimental Section

*Chemicals and Materials:*

Peroxidase from horseradish type VI (HRP) (E.C. 1.11.1.7), hydrogen peroxide, 2-methoxyphenol (guaiacol), potassium hexacyanoferrate (II) trihydrate (ferrocyanide; $K_4Fe(CN)_6$), β-nicotinamide adenine dinucleotide reduced dipotassium salt (NADH), L-ascorbic acid (Asc) and L-glutathione reduced (GSH) were purchased from Sigma-Aldrich. 2,2'-azino-*bis*(3-ethylbenzothiazoline-6-sulfonic acid) (ABTS) was purchased from Fluka. All chemicals were used as supplied without further purification. Ultrapure water (18.2 MΩ·cm) from a NANOpure Diamond (Barnstead) source was used in all of the experiments. All measurements were performed in 0.01 M phosphate buffer, pH = 6.0, at a temperature of $23 \pm 2$ °C.

*Method*:

For each system, we took $H_2O_2$ and a chromogen: ABTS, guaiacol, or ferrocyanide, as the logic inputs. The chemical reaction was catalyzed by HRP and the filtering effect was accomplished with the added Asc, NADH, or GSH. The logic output was the oxidized chromogen, the concentration of which was measured by monitoring the optical absorbance, Abs, at λ = 415 nm, which is near the centers of broad spectral peaks distinguishing all three dyes from their original non-oxidized chromogens, as a function of time, $t$. The biocatalytic reactions were performed in a 96-well microtiter plate (VWR) and the absorbance measurements were carried out using a Model 680 (Bio-Rad Laboratories) microplate reader.

**3. Description of the System and Its Modeling**

Our realization of the **AND** gate is rather standard in the context of biocatalytic reactions interpreted as binary logic functions.[12] The two inputs were $H_2O_2$ and one of the three chromogens: ABTS, ferrocyanide, or guaiacol. The latter case is illustrated in Figure 1, which also offers a schematic of the processes involved. The "signal processing" was biocatalyzed by HRP. The output was measured optically as the amount of the oxidized chromogen. This system was in fact already studied as the AND gate with the comparison of ABTS and ferrocyanide in certain kinetic regimes.[50] Here the regime of the gate operation is different, and we use the new "filtering" approach to investigate the possibility of making the gate double-sigmoid. Specifically, we use the earlier considered[55] Asc as the filter-process reducing agent, but we also check NADH and GSH in the favorable configuration (realized with guaiacol as the chromogen; see Figure 1) once it is identified (see Section 4). The initial concentrations of the inputs can vary, but two selected (as appropriate for application) values are considered as logic **0** and **1**. Some of the chemicals are considered the "gate machinery" in that their initial concentrations can be chosen to achieve the desired functioning. Here these are HRP (having the same value in all the experiments) and the reducing agent. The logic **0** and **1** of the output at a selected "gate time" $t_g > 0$ are set by the gate function itself. In the present case we started with or adjusted various parameters at experimentally convenient values, largely based on earlier experience with such systems. The logic **0** values were set at physical zeros.

In applications,[40,41,51] biocatalytic "gates" function in environments with noisy inputs. Furthermore, the gate function's actual chemical realization can also add to the noise in the output. This makes it difficult to use such logic elements as units of larger networks. Optimization of the system parameters[31,49] can decrease the amplification of the noise. A better, recent approach involves the added "filtering," which changes the typically convex shape of the response surface into sigmoid.[51-56] In our case, the reducing agent "recycles" a fraction of the output, converting it back into the original chromogen. With the proper selection of the parameters, the response will generally become sigmoid in the input $H_2O_2$, without a significant loss of the output signal intensity. However, in order to achieve effective noise suppression[51-56] in the vicinity of all the four logic-input combinations, we actually seek a response with a double-sigmoid shape. The extent to which this can be accomplished with the present **AND** gate system is studied here. As mentioned, we are also interested in the extent of "modularity" of this system, as the choices of the input chromogen and reducing agent are varied as part of the gate-function optimization. More generally, modularity refers to the degree to which a multiunit system's functional units may be adjusted. The concept of modularity in Biology, for instance, relates to the property that metabolic pathways are composed of well-defined interconnected units of various complexity.[65-67]

Let us outline our modeling approach, which consists of two steps. First, we fit experimental data (as illustrated in Figure 1) in terms of the actual physical values measured. Second, the fitted "response surface" is used to analyze the gate's performance. This is done in terms of the scaled "logic range" variables,[49,50]

$$x = [H_2O_2](0)/[H_2O_2]_1 , \qquad (1)$$

$$y = [D](0)/[D]_1 , \qquad (2)$$

$$z = [D_{ox}](t_g)/[D_{ox}]_{1,1} . \qquad (3)$$

Here $D$ denotes the chromogen, $D_{ox}$ in its oxidized form (the output), and the values at the logic **1** (shown by subscripts) for the inputs are those at time 0, whereas for the output, $[D_{ox}]_{1,1}$ is the value obtained for both inputs at the logic **1,1**, at time $t_g$. If the function $z(x,y)$ is smooth (as is the case here), then a simple estimate of the noise transmission factor of the gate in the vicinity of each logic point is obtained by calculating the slope, i.e., the absolute value of the gradient, $|\vec{\nabla}z(x,y)|$, evaluated at and near $(x,y) = (0,0), (0,1), (1,0)$, and $(1,1)$. This assumes that the fluctuations in the input signals due to noise are symmetrical in terms of the two inputs once these are scaled to the logic-variable ranges between 0 and 1. Filtering aims to make the values of the noise transmission factor smaller than 1 in the $(x,y)$ range near each logic point which exceeds the noise spread in the inputs in terms of $(x,y)$. Of course, transmission of noise from the inputs to output is not the only source of the possible noise in the output. Other sources are possible, the most obvious one is the inaccuracy of the experimental realization of (i.e., the noise in the values of) the function $z(x,y)$. We further comment on this matter in Section 4. However, for large-scale networking it has been argued[31,68] that the important source of noise to eliminate in as many "gates" as possible at the level of single-gate optimization, is noise amplification.

The mechanism of the biocatalytic action of HRP has many possible pathways[69-71] which involve numerous rate constants. These rate constants in turn depend on the chemical and physical conditions in the system, notably, the pH and temperature. However, given the limited quality of typically available experimental data and the fact that we only need a qualitative description of the overall shape of the response surface, for logic-gate design it has been argued that we can use a simplified, few-parameter kinetic model[31,49] or other even more phenomenological shape-fitting models.[56] Here we use the following approximate Michaelis-Menten type kinetic description of the kinetics in the primary pathway of HRP,

$$E + H_2O_2 \underset{k_{-1}}{\overset{k_1}{\rightleftarrows}} C, \tag{4}$$

$$C + D \overset{k_2}{\rightarrow} E + D_{ox}, \tag{5}$$

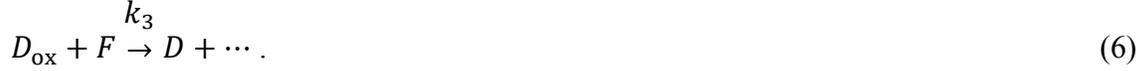

$$D_{ox} + F \xrightarrow{k_3} D + \cdots . \tag{6}$$

Here the last step is the added irreversible filtering process, and we use the following notation for the (time-dependent) chemical concentrations: $E$ stands for the enzyme and $C$ for a complex formed in the first step of the biocatalytic process, $D$ and $D_{ox}$ denote the chromogen and the oxidized (dye) form, respectively, and $F$ is the filtering compound. Information on the initial concentration ranges of various chemicals is given in Table 1, with the initial concentration of the enzyme always set to $E_0 = E(0) = 4.5 \times 10^{-3}$ μM. The output values are measured via the absorbance at the gate times, $t_g$, of 15 min (no filtering) or 30 min (with added filtering), as explained in Section 4.

In order to reduce the number of adjustable parameters in data fitting with the reaction scheme of Equations (4)-(6), we make several simplifying assumptions following a standard practice in the logic-gate modeling of this type of systems,[55] based on the experimentally known properties of peroxidases in the considered regime of kinetics driven by supply of inputs.[69-73] We set $k_{-1} = 0$, ignoring the back-reaction in Equation (4).[55,73] Furthermore, the process shown in Equation (5) actually involves a formation of a second intermediate complex as part of the biocatalytic functioning of HRP,[69-71] which for our purposes can be represented by a single effective irreversible step as shown.[72] We then fit data without filtering to find the values of $k_1$ and $k_2$ for each particular system, and then fit a single added parameter, $k_3$, by using data with filtering. Our results for $k_1$ and $k_2$ presented in Section 4, are approximately consistent with published data for a related system involving the biocatalytic activity of HRP.[73] The rate equations for the considered processes are

$$\frac{dH}{dt} = -k_1 H(E_0 - C), \tag{7}$$

$$\frac{dC}{dt} = k_1 H(E_0 - C) - k_2 C(D_0 - P), \tag{8}$$

$$\frac{dP}{dt} = k_2 C(D_0 - P) - k_3 FP , \qquad (9)$$

$$\frac{dF}{dt} = -k_3 FP , \qquad (10)$$

where $H = [H_2O_2](t)$, $E_0 = [HRP](0)$, $D_0 = [D](0)$, and the final product is denoted as $P = [D_{ox}](t)$. All the other notations were defined earlier, and we also used the relations

$$E(t) = E_0 - C(t) , \qquad (11)$$

$$D(t) = D_0 - P(t) . \qquad (12)$$

The initial conditions for Equations (7)-(10) were as follows. For the inputs, $H(0)$ and $D_0$, we use the initial concentrations of $H_2O_2$ and the proper chromogen, varied in the range between the logic **0** (here zero concentrations) and the logic **1** values given in Table 1. The specific logic **1** values (and in principle also logic **0** values) for the inputs are set by the gate's applications/networking. For our model study we selected representative values for logic **1**, based on earlier work, with vastly varying ranges for $H_2O_2$, as commented on later. Note that the gate function itself sets the logic values of its output. When the response surface is not smooth, values outside the logic range (for instance, somewhat beyond the logic **1**) can also be sampled for better mapping, but here no such data were taken. For the "gate machinery" chemicals, the initial concentration of HRP was already mentioned, $4.5 \times 10^{-3}$ µM, whereas those of the filtering chemical reducing agents used, are listed in Table 1.

### 4. Results and Discussion

Let us first consider the system without the filtering process, i.e., $F(0) = 0$. The values of the two adjustable parameters, $k_1$ and $k_2$, depend on the chromogen used. In Figure 2 we illustrate the data (and the fitting surface) for the three chromogens considered, generally measured as a function of time but here shown for a convenient "gate time" of $t_g$ = 15 min, for which we get the typical enzymatic-gate response surfaces which are convex and show saturation for large enough inputs. The fitted rate constant values were $(k_1, k_2)$ = (15.43, 3.48), (0.98, 0.21), (8.79×10$^{-2}$, 2.59×10$^{-2}$) s$^{-1}$μM$^{-1}$, for the gates with ABTS, guaiacol, ferrocyanide, respectively. The surfaces in Figure 2 were used to conclude that the studied **AND** gates amplify noise by factors of 4.8 for ABTS, 7.6 for guaiacol, and 4.4 for ferrocyanide, with the largest slope (among the slopes at the four logic points) realized at inputs **10**, i.e., for maximum H$_2$O$_2$ concentration and zero chromogen concentration, in all three cases. Recall that for these relatively smooth gate-response surfaces we can use the slope,[68]

$$|\vec{\nabla}z(x,y)| = \sqrt{\left(\frac{\partial z}{\partial x}\right)^2 + \left(\frac{\partial z}{\partial y}\right)^2}, \tag{13}$$

the largest value of which among the four logic points provides the estimate of the noise noise-transmission factor: amplification, if larger than 1, or suppression, if lower than 1.

The convex shape of such a response can be modified by varying some of the parameters of the enzymatic process, as was extensively studied in earlier works on gate optimization[31,49] aimed at minimizing noise amplification. However, gate optimization can at best decrease the amplification effect, rather than yield noise suppression. With regards to modularity, it is demonstrated here only to the extent that, with the identical "gate machinery" and gate time, reasonable convex-response surfaces with comparable signal intensity can be obtained for vastly differing ranges of the primary input, H$_2$O$_2$ (see Figure 2 and Table 1), provided we properly select the secondary input (chromogen).

Earlier studies found that ascorbate, which is an efficient and commonly used reducing agent, can efficiently reverse the oxidation by H$_2$O$_2$, biocatalytically enabled by HRP, thus acting as a "filtering chemical" with respect to the oxidizing agent (H$_2$O$_2$), the the latter as the primary input of the **AND** gate. Specifically, this was demonstrated[55] with 3,5,3',5'-tetramethylbenzidine (TMB), which has more complicated kinetics,[72] as the chromogen being oxidized. As mentioned, here we aim at exploring the extent to which such an added filtering involving "recycling" of the secondary input (chromogen) can yield a double-sigmoid response. We find that while recycling prevents a significant drop in the output signal intensity, it also precludes effective filtering with respect to the input into which output is being recycled.

Figure 3 illustrates our three **AND** gates in a situation in which the initial concentrations of ascorbate were large enough to result in a relatively intense filtering with respect to $H_2O_2$ as an input. Indeed, the output signal was practically eliminated for sizable (on the scale of the logic ranges) fractions of the input $H_2O_2$ values. However, only a very small, for ABTS, or barely noticeable, for $K_4Fe(CN)_6$, filtering effect was obtained along the chromogen-concentration axis, with a moderate effect for guaiacol, for which the response is generally more symmetrical. The price paid for this "over-filtering" is a possible shrinkage of a well-defined saturated regime near the logic-**1** output value, as well as some loss of intensity. The latter effect was largely avoided by doubling the gate time for the filtered systems. We note that here the model surfaces shown for the "filtered" systems represent the fitting involving a single rate constant, $k_3$, with the other two rate constants taken from the earlier fitting of the non-filtered data, according to the rate equations introduced in Section 3. The values fitted for gates with ABTS, guaiacol, $K_4Fe(CN)_6$ were, respectively, $k_3 = 1.76\times10^{-1}, 1.4\times10^{-3}, 2.6\times10^{-4}$ $s^{-1}\mu M^{-1}$.

We note that the sigmoid behavior, which was obtained symmetrically in both inputs only for the case of guaiacol, clearly seen in the middle panel in Figure 3 (and also in Figure 1), is obtained with respect to the variation of the inputs at a fixed gate time. The time-dependence of the product concentration (output signal) in enzymatic reactions always has a transient nonlinear regime, usually short, followed by an approximately linear regime and then saturation. Added "filtering," which is this case involves the "recycling" of a part of the product, can significantly extend this transient regime of the time dependence as well, but this is not the effect that we are interested in when using the system as a "logic gate."

The "filtering" properties are further explored in Figure 4, which shows the values of the slope, $|\vec{\nabla}z(x,y)|$, for the three realized "over-filtered" systems. A good-quality double-sigmoid response should have substantial and symmetrically positioned regions with slope less than 1, near each of the logic points, for input-noise suppression on its transmission to the output. For guaiacol, the "over-filtering" primarily resulted in such a "flat response region" near input **11** being much smaller than "flat regions" near inputs **00**, **01**, **10**. ABTS offers an example of a system for which even with "over-filtering," which notably shrank the flat region near logic-**1** in terms of response to $H_2O_2$, only a very small region of flat response could be obtained along the ABTS input direction, specifically, near inputs **10**. Finally, for the $K_4Fe(CN)_6$ system, no flat region was obtained at all near inputs **10**. Thus, the latter system actually amplifies noise at inputs **10**, by a factor of about 3.2. The shown ABTS and guaiacol **AND** gates can tolerate (suppress) noise of order 7 % and 18 %, respectively, as percentages of the logic ranges of the inputs. This was estimated by the closest distance that the lines marking the value of the slope $|\vec{\nabla}z(x,y)| = 1$, see Figure 4, approach the logic points.

Our modeling approach is approximate, but we expect it to have a reasonable extrapolative value to allow optimization of the gate functioning. Specifically, "over-filtering" resulted in the inflection region (of large slope) of the double-sigmoid response, when obtained, being not symmetrically positioned. For the favorable case of guaiacol, we then considered a parameter modification to make the double-sigmoid more symmetrical. Figure 5 shows the data measured with the reduced filtering chemical, ascorbate, concentration of 120 µM, selected as suggested by the model. The model surface presented in Figure 5, however, was not fitted to these data but calculated with the earlier-determined rate constants, and shows a good agreement with the experiment. The resulting inflection region (of slope larger than 1) bounded by the two black solid lines in the slope mapping plot in the figure, is now relatively symmetrically positioned.

The noise tolerance (the smallest fractional distance of this region from all the logic points) for this gate is 16 %, which is actually not an improvement over the asymmetrical case. However, this exercise demonstrates an interesting property of the filtering approach. Since enzymatic logic gates of this type usually have large scaled-variable slopes at low inputs (where the initially linear response bends to become convex to match saturation at larger inputs), most parameter modifications to improve or otherwise change the gate's behavior have to affect the small-inputs regime. We can contrast two "gate machinery" chemicals. The initial concentration of the enzyme, $E_0$, if adjusted, affects the reaction output approximately linearly in this small-inputs regime, and therefore its variation cancels out in the leading order, in terms of the scaled (to the logic ranges) variables entering the function of interest, $z(x,y)$. Therefore, large, order-of-magnitude changes in this and other "internal" gate machinery parameters are required for any significant modifications of the gate-response function. This was noted and identified as a problem in earlier studies.[31,74] With the externally added filtering, however, this problem is not present, because filtering affects the low-input regime directly, and therefore adjustments of "external" gate-machinery parameters, here, the initial concentration of the reducing agent, $F(0) = [\text{Asc}](0)$, by reasonable amounts suffice for achieving noticeable change in the gate response function.

Next, we tested the "modularity" of the gate functioning in terms of the use of different reducing agents (filtering chemicals). Ascorbate was replaced with NADH and also with GSH, in both cases with the same initial concentration, $F(0) = 120$ μM, as the just considered Asc case (Figure 5). These results are shown in Figure 6 (for NADH) and Figure 7 (for GSH). The model surfaces were obtained by fitting the rate constant $k_3$, with the other two rate constants taken from the unfiltered data as before. The values fitted for filtering with NADH, GSH were, respectively, $k_3 = 8.2 \times 10^{-3}, 12.3 \times 10^{-3}$ s$^{-1}$μM$^{-1}$. The slopes (noise transmission factors) are also mapped out in the figures. We conclude that replacement of the reducing agent while keeping all the other parameters unchanged, is not entirely modular because in both cases it resulted in "over-filtering" as compared to the Asc case. This is obviously related to the fact that both these reducing agents are more active than ascorbate in the present guaiacol system, for which the value of $k_3$ reported earlier in this section was only $1.4 \times 10^{-3}$ s$^{-1}$μM$^{-1}$.

It is, however, useful to look at the present set of illustrative systems to address the following interesting matter. Noise amplification during transmission from inputs to output is only one source of the overall noise buildup in systems involving separate "gates." While its suppression is paramount for scalability[31] in network design, other sources of noise should be avoided or minimized as much as possible as part of the overall optimization of biocomputing gates. All experimental gate-function surfaces shown demonstrate that imprecise gate-response realization constitutes another important source of noise. Biochemical data are always relatively noisy at each step of processing. The actual data for the three filtered systems shown in Figure 3, for instance, are spread approximately 5.0 %, 3.5 % and 4.5 % (measured as a root-mean-squared displacement from the fitted surface) for ABTS, guaiacol, $K_4Fe(CN)_6$ as the chromogen, respectively, which should be compared to the earlier-reported error tolerances of 7 % for ABTS, and 18 % for guaiacol, whereas the $K_4Fe(CN)_6$ is noise-amplifying (technically, zero noise tolerance at the gate level). This indicates that the ABTS and guaiacol gates at least, do not "tax" the network by adding extra noise as long as it is kept within several percentage points on the relative scale of the signal spans. The second guaiacol gate, shown in Figure 5, has a somewhat larger gate-realization noise level, approximately 8 %, but its noise tolerance of 16 % indicates that it is also usable as a noise-tolerant/noise-suppressing single-step element in a network. However, the gates obtained for guaiacol with the other two reducing agents, as shown in Figure 6 (for NADH) and Figure 7 (for GSH), have gate-realization added-noise levels of about 10 % for NADH, and 13 % for GSH. The locations of the black solid lines in the slope mappings shown in Figures 6 and 7 indicate that the input-noise tolerances of these gates are, respectively, 9 % and 10 %, which suggests that the use of faster reducing agents in this case, with all the other parameters equal, made the gates overall less "noise-handling friendly" in a network setting.

## 5. Conclusions

Design ideas for gates of the type explored here have been of interest in many biocomputing and biosensing applications. We find that the concept of modularity in the sense of replacing specific chemicals according to the operational regime, signal ranges and their nature, and other constraints for such gates is applicable but not in a straightforward manner. Respective adjustments of other parameters will likely be needed, depending on the gate's application and the noise levels in its environment, in order to ensure that it is networkable.

While generally "recycling" as part of the filtering prevents a significant loss of the overall signal intensity, our filtered data shown in the figures illustrate that this is another quantity that is not systematically predictable. The intensity is reduced in various degrees, and this contributes to having a larger relative level of noise added by the gate realization, as discussed in Section 4.

Thus, there is no simple prescription for making the resulting gates symmetrically double-sigmoid, and well-behaved as far as the overall noise handling is concerned. However, with a certain degree of experimental tinkering, including the selection of proper chemicals, and the use of the developed theoretical analysis approach the best systems can be identified for specific applications of interest.

We note that this is the first systematic study of design ideas for two-input gates with regards to devising an efficient gate-modification approach involving a simple added chemical process and aimed at a good-quality double-sigmoid response. In future studies it would be interesting to attempt networking of such gates as well as compare them to those involving different chemical or physical modifications, such as consumption/deactivation of fractions of both inputs, or a fraction of the output.

## 6. Acknowledgements

We acknowledge support of this research by the National Science Foundation under grants CBET-1066397 and CCF-1015983.

**REFERENCES**

(1) *Molecular and Supramolecular Information Processing – From Molecular Switches to Unconventional Computing.* Katz, E. (Ed.), Willey-VCH, Weinheim, **2012**.

(2) De Silva, A. P.; Uchiyama, S.; Vance, T. P.; Wannalerse, B. A Supramolecular Chemistry Basis for Molecular Logic and Computation. *Coord. Chem. Rev.* **2007**, *251*, 1623–1632.

(3) Szacilowski, K. Digital Information Processing in Molecular Systems. *Chem. Rev.* **2008**, *108*, 3481–3548.

(4) Credi, A. Molecules that Make Decisions. *Angew. Chem. Int. Ed.* **2007**, *46*, 5472–5475.

(5) Pischel, U. Chemical Approaches to Molecular Logic Elements for Addition and Subtraction. *Angew. Chem. Int. Ed.* **2007**, *46*, 4026–4040.

(6) Andreasson, J.; Pischel, U. Smart Molecules at Work – Mimicking Advanced Logic Operations. *Chem. Soc. Rev.* **2010**, *39*, 174–188.

(7) *Biomolecular Computing – From Logic Systems to Smart Sensors and Actuators.* Katz, E. (Ed.), Willey-VCH, Weinheim, **2012**.

(8) Shapiro, E.; Gil, B. Biotechnology − Logic Goes in Vitro. *Nat. Nanotechnol.* **2007**, *2*, 84–85.

(9) Benenson, Y. Biocomputers: From Test Tubes to Live Cells. *Mol. Biosyst.* **2009**, *5*, 675–685.

(10) Ashkenasy, G.; Ghadiri, M. R. Boolean Logic Functions of a Synthetic Peptide Network. *J. Am. Chem. Soc.* **2004**, *126*, 11140–11141.

(11) Stojanovic, M. N.; Stefanovic, D.; LaBean, T.; Yan, H. Computing with Nucleic Acids. In: *Bioelectronics: From Theory to Applications*, Willner, I.; Katz, E. (Eds.) Wiley-VCH, Weinheim, **2005**, pp. 427–455.

(12) Katz, E.; Privman, V. Enzyme-Based Logic Systems for Information Processing. *Chem. Soc. Rev.* **2010**, *39*, 1835–1857.

(13) *Unconventional Computation. Lecture Notes in Computer Science*, Calude, C. S.; Costa, J. F.; Dershowitz, N.; Freire, E.; Rozenberg, G. (Eds.), Vol. 5715, Springer, Berlin, **2009**.

(14) *Unconventional Computing*, Adamatzky, A.; De Lacy Costello, B.; Bull, L.; Stepney, S.; Teuscher, C. (Eds.), Luniver Press, UK, **2007**.

(15) de Murieta, I. S.; Miro-Bueno, J. M.; Rodriguez-Paton, A. Biomolecular Computers. *Curr. Bioinformatics* **2011**, *6*, 173–184.

(16) Wang, J.; Katz, E. Digital Biosensors with Built-in Logic for Biomedical Applications – Biosensors Based on Biocomputing Concept. *Anal. Bioanal. Chem.* **2010**, *398*, 1591–1603.

(17) Wang, J.; Katz, E. Digital Biosensors with Built-in Logic for Biomedical Applications. *Isr. J. Chem.* **2011**, *51*, 141–150.

(18) Katz, E.; Wang, J.; Privman, M.; Halámek, J. Multianalyte Digital Enzyme Biosensors with Built-in Boolean Logic. *Anal. Chem.* **2012**, *84*, 5463–5469.

(19) Stojanovic, M. N. Some Experiments and Directions in Molecular Computing and Robotics. *Isr. J. Chem.* **2011**, *51*, 99–105.

(20) Pei, R.; Matamoros, E.; Liu, M.; Stefanovic, D.; Stojanovic, M. N. Training a Molecular Automaton to Play a Game. *Nat. Nanotechnol.* **2010**, *5*, 773–777.

(21) Privman, V. Biomolecular Computing: Learning Through Play. *Nat. Nanotechnol.* **2010**, *5*, 767–768.

(22) Kahan, M.; Gil, B.; Adar, R.; Shapiro, E. Towards Molecular Computers that Operate in a Biological Environment. *Physica D* **2008**, *237*, 1165–1172.

(23) Babaei, M. A Novel Text and Image Encryption Method Based on Chaos Theory and DNA Computing. *Natural Computing* **2013**, *12*, 101–107.

(24) Domanskyi, S.; Privman, V. Design of Digital Response in Enzyme-Based Bioanalytical Systems for Information Processing Applications. *J. Phys. Chem. B* **2012**, *116*, 13690–13695.

(25) Katz, E.; Minko, S.; Halámek, J.; MacVittie, K.; Yancey, K. Electrode Interfaces Switchable by Physical and Chemical Signals for Biosensing, Biofuel and Biocomputing Applications. *Anal. Bioanal. Chem.* **2013**, *405*, 3659–3672.

(26) Katz, E.; Bocharova, V.; Privman, M. Electronic Interfaces Switchable by Logically Processed Multiple Biochemical and Physiological Signals. *J. Mater. Chem.* **2012**, *22*, 8171–8178.

(27) Bocharova, V.; Katz, E. Switchable Electrode Interfaces Controlled by Physical, Chemical and Biological Signals. *Chemical Record* **2012**, *12*, 114–130.

(28) Katz, E. Bioelectronic Devices Controlled by Biocomputing Systems. *Isr. J. Chem.* **2011**, *51*, 132–140.

(29) Privman, M.; Tam, T. K.; Pita, M.; Katz, E. Switchable Electrode Controlled by Enzyme Logic Network System: Approaching Physiologically Regulated Bioelectronics. *J. Am. Chem. Soc.* **2009**, *131*, 1314–1321.

(30) Krämer, M.; Pita, M.; Zhou, J.; Ornatska, M.; Poghossian, A.; Schöning, M. J.; Katz, E. Coupling of Biocomputing Systems with Electronic Chips: Electronic Interface for Transduction of Biochemical Information. *J. Phys. Chem. C* **2009**, *113*, 2573–2579.

(31) Privman, V.; Arugula, M. A.; Halámek, J.; Pita, M.; Katz, E. Network Analysis of Biochemical Logic for Noise Reduction and Stability: A System of Three Coupled Enzymatic AND Gates. *J. Phys. Chem. B* **2009**, *113*, 5301–5310.

(32) Niazov, T.; Baron, R.; Katz, E.; Lioubashevski, O.; Willner, I. Concatenated Logic Gates Using Four Coupled Biocatalysts Operating in Series. *Proc. Natl. Acad. USA.* **2006**, *103*, 17160–17163.

(33) Baron, R.; Lioubashevski, O.; Katz, E.; Niazov, T.; Willner, I. Elementary Arithmetic Operations by Enzymes: A Model for Metabolic Pathway Based Computing. *Angew. Chem. Int. Ed.* **2006**, *45*, 1572–1576.

(34) Baron, R.; Lioubashevski, O.; Katz, E.; Niazov, T.; Willner, I. Logic Gates and Elementary Computing by Enzymes. *J. Phys. Chem. A* **2006**, *110*, 8548–8553.

(35) May, E. E.; Dolan, P. L.; Crozier, P. S.; Brozik, S.; Manginell, M. Towards De Novo Design of Deoxyribozyme Biosensors for GMO Detection. *IEEE Sens. J.* **2008**, *8*, 1011–1019.

(36) von Maltzahn, G.; Harris, T. J.; Park, J.-H.; Min, D.-H.; Schmidt, A. J.; Sailor, M. J.; Bhatia, S. N. Nanoparticle Self-Assembly Gated by Logical Proteolytic Triggers. *J. Am. Chem. Soc.* **2007**, *129*, 6064–6065.
– 18 –

(37) Gil, B.; Kahan-Hanum, M.; Skirtenko, N.; Adar, R.; Shapiro, E. Detection of Multiple Disease Indicators by an Autonomous Biomolecular Computer. *Nano Lett.* **2011**, *11*, 2989–2996.

(38) Halámková, L.; Halámek, J.; Bocharova, V.; Wolf, S.; Mulier, K. E.; Beilman, G.; Wang, J.; Katz, E. Analysis of Biomarkers Characteristic of Porcine Liver Injury – From Biomolecular Logic Gates to Animal Model. *Analyst* **2012**, *137*, 1768–1770.

(39) Zhou, N.; Windmiller, J. R.; Valdés Ramírez, G.; Zhou, M.; Halámek, J.; Katz, E.; Wang, J. Enzyme-Based NAND Gate for Rapid Electrochemical Screening of Traumatic Brain Injury in Serum. *Anal. Chim. Acta* **2011**, *703*, 94–100.

(40) Zhou, J.; Halámek, J.; Bocharova, V.; Wang, J.; Katz, E. Bio-Logic Analysis of Injury Biomarker Patterns in Human Serum Samples. *Talanta* **2011**, *83*, 955–959.

(41) Halámek, J.; Bocharova, V.; Chinnapareddy, S.; Windmiller, J. R.; Strack, G.; Chuang, M.-C.; Zhou, J.; Santhosh, P.; Ramirez, G. V. Arugula, M. A.; et al. Multi-Enzyme Logic Network Architectures for Assessing Injuries: Digital Processing of Biomarkers. *Mol. Biosyst.* **2010**, *6*, 2554–2560.

(42) Unger, R.; Moult, J. Towards Computing with Proteins. *Proteins* **2006**, *63*, 53–64.

(43) Stojanovic, M. N.; Stefanovic, D. Chemistry at a Higher Level of Abstraction. *J. Comput. Theor. Nanosci.* **2011**, *8*, 434–440.

(44) Ezziane, Z. DNA Computing: Applications and Challenges. *Nanotechnology*, **2006**, *17*, R27–R39.

(45) Rinaudo, K.; Bleris, L.; Maddamsetti, R.; Subramanian, S.; Weiss, R.; Benenson, Y. A Universal RNAi-Based Logic Evaluator that Operates in Mammalian Cells. *Nat. Biotechnol.* **2007**, *25*, 795–801.

(46) Tamsir, A.; Tabor, J. J.; Voigt, C. A. Robust Multicellular Computing Using Genetically Encoded NOR Gates and Chemical 'Wires'. *Nature* **2011**, *469*, 212–215.

(47) Li, Z.; Rosenbaum, M. A.; Venkataraman, A.; Tam, T. K.; Katz, E.; Angenent, L. T. Bacteria-Based AND Logic Gate: A Decision-Making and Self-Powered Biosensor. *Chem. Commun.* **2011**, *47*, 3060–3062.

(48) Arugula, M. A.; Shroff, N.; Katz, E.; He, Z. Molecular AND Logic Gate Based on Bacterial Anaerobic Respiration. *Chem. Commun.* **2012**, *48*, 10174–10176.

(49) Privman, V.; Strack, G.; Solenov, D.; Pita, M.; Katz, E. Optimization of Enzymatic Biochemical Logic for Noise Reduction and Scalability: How Many Biocomputing Gates Can Be Interconnected in a Circuit? *J. Phys. Chem. B* **2008**, *112*, 11777–11784.

(50) Melnikov, D.; Strack, G.; Pita, M.; Privman, V.; Katz, E. Analog Noise Reduction in Enzymatic Logic Gates. *J. Phys. Chem. B* **2009**, *113,* 10472–10479.

(51) Halámek, J.; Zavalov, O.; Halámková, L.; Korkmaz, S.; Privman, V.; Katz, E. Enzyme-Based Logic Analysis of Biomarkers at Physiological Concentrations: AND Gate with Double-Sigmoid "Filter" Response. *J. Phys. Chem. B* **2012**, *116*, 4457–4464.

(52) Zavalov, O.; Bocharova, V.; Privman, V.; Katz, E. Enzyme-Based Logic: OR Gate with Double-Sigmoid Filter Response. *J. Phys. Chem. B* **2012**, *116*, 9683–9689.

(53) Zavalov, O.; Bocharova, V.; Halámek, J.; Halámková, L.; Korkmaz, S.; Arugula, M. A.; Chinnapareddy, S.; Katz, E.; Privman, V. Two-Input Enzymatic Logic Gates Made Sigmoid by Modifications of the Biocatalytic Reaction Cascades. *Int. J. Unconv. Comput.* **2012**, *8*, 347–365.

(54) Pita, M.; Privman, V.; Arugula, M. A.; Melnikov, D.; Bocharova, V.; Katz, E. Towards Biochemical Filter with Sigmoidal Response to pH Changes: Buffered Biocatalytic Signal Transduction. *Phys. Chem. Chem. Phys.* **2011**, *13*, 4507–4513.

(55) Privman, V.; Halámek, J.; Arugula, M. A.; Melnikov, D.; Bocharova, V.; Katz, E. Biochemical Filter with Sigmoidal Response: Increasing the Complexity of Biomolecular Logic. *J. Phys. Chem. B* **2010**, *114*, 14103–14109.

(56) Privman, V.; Fratto, B. E.; Zavalov, O.; Halámek, J.; Katz, E. Enzymatic AND Logic Gate with Sigmoid Response Induced by Photochemically Controlled Oxidation of the Output. *J. Phys. Chem. B* **2013**, *117*, 7559–7568.

(57) Rafael, S. P.; Vallée-Bélisle, A.; Fabregas, E.; Plaxco, K.; Palleschi, G.; Ricci, F. Employing the Metabolic "Branch Point Effect" to Generate an All-or-None,

Digital-Like Response in Enzymatic Outputs and Enzyme-Based Sensors. *Anal. Chem.* **2012**, *84*, 1076–1082.

(58) Vallée-Bélisle, A.; Ricci, F.; Plaxco, K. W. Engineering Biosensors with Extended, Narrowed, or Arbitrarily Edited Dynamic Range. *J. Am. Chem. Soc.* **2012**, *134*, 2876−2879.

(59) Kang, D.; Vallée-Bélisle, A.; Plaxco, K. W.; Ricci, F. Re-engineering Electrochemical Biosensors To Narrow or Extend Their Useful Dynamic Range. *Angew. Chem. Int. Ed.* **2012**, *51*, 6717–6721.

(60) Ricci, F.; Vallée-Bélisle, A.; Plaxco, K. W. High-Precision, In Vitro Validation of the Sequestration Mechanism for Generating Ultrasensitive Dose-Response Curves in Regulatory Networks. *PLoS Comput. Biol.* **2011**, *7*, article #e1002171.

(61) Privman, V.; Pedrosa, V.; Melnikov, D.; Pita, M.; Simonian, A.; Katz, E. Enzymatic AND-Gate Based on Electrode-Immobilized Glucose-6-phosphate Dehydrogenase: Towards Digital Biosensors and Biochemical Logic Systems with Low Noise. *Biosens. Bioelectron.* **2009**, *25*, 695–701.

(62) Pedrosa, V.; Melnikov, D.; Pita, M.; Halámek, J.; Privman, V.; Simonian, A.; Katz, E. Enzymatic Logic Gates with Noise-Reducing Sigmoid Response. *Int. J. Unconv. Comput.* **2010**, *6*, 451–460.

(63) Qian, H.; Shi, P.-Z. Fluctuating Enzyme and Its Biological Functions: Positive Cooperativity without Multiple States. *J. Phys. Chem. B* **2009**, *113*, 2225–2230.

(64) Rabinowitz, J. D.; Hsiao, J. J.; Gryncel, K. R.; Kantrowitz, E. R.; Feng, X.-J. Dissecting Enzyme Regulation by Multiple Allosteric Effectors: Nucleotide Regulation of Aspartate Transcarbamoylase. *Biochemistry* **2008**, *47*, 5881–5888.

(65) Ravasz, E.; Somera, A. L.; Mongru, D. A.; Oltvai, Z. N.; Barabasi, A. L. Hierarchical Organization of Modularity in Metabolic Networks. *Science* **2002**, *297*, 1551–1555.

(66) Sridharan, G. V.; Hassoun, S.; Lee, K. Identification of Biochemical Network Modules Based on Shortest Retroactive Distances. *PLoS Comput. Biol*. **2011**, *7*, article #e1002262.

(67) Clune, J.; Mouret, J-B.; Lipson, H. The Evolutionary Origins of Modularity *Proc. R. Soc. B* **2013**, *280*, article #20122863.

(68) Privman, V. Error-Control and Digitalization Concepts for Chemical and Biomolecular Information Processing Systems. *J. Comput. Theor. Nanosci.* **2011**, *8*, 490–502.

(69) Dunford, H. B. Horseradish Peroxidase: Structure and Kinetic Properties. In: *Peroxidases in Chemistry and Biology*, Everse, J.; Everse, K. E.; Grisham, M. B. (Eds.). Vol. 2. CRC Press, Boca Raton, Florida, **1991**, pp. 1–24.

(70) Dunford, H. B. *Heme Peroxidases*. Wiley–VCH, New York, **1999**.

(71) Veitch, N. C.; Smith, A. T. Horseradish Peroxidase. *Adv. Inorg. Chem.* **2001**, *51*, 107–162.

(72) Josephy, P. D.; Eling, T.; Mason, R. P. The Horseradish Peroxidase-Catalyzed Oxidation of 3,5,3',5'-Tetramethylbenzidine. Free Radical and Charge-Transfer Complex Intermediates. *J. Biol. Chem.* **1982**, *257*, 3669–3675.

(73) Marquez, L. A.; Dunford, H. B. Mechanism of the Oxidation of 3,5,3',5'-Tetramethylbenzidine by Myeloperoxidase Determined by Transient- and Steady-State Kinetics. *Biochemistry* **1997**, *36*, 9349–9355.

(74) Arugula, M. A.; Halámek, J.; Katz, E.; Melnikov, D.; Pita, M.; Privman, V.; Strack, G. Optimization of Enzymatic Logic Gates and Networks for Noise Reduction and Stability. In: *Proc. Conf. CENICS 2009*, Kent, K. B.; Dini, P.; Franza, O.; Palacios, T.; Reig, C.; Maranon, J. E.; Rostami, A.; Zammit-Mangion, D.; Hasenplaugh, W. C.; Toran F.; Zafar Y. (Eds.). IEEE Comp. Soc. Conf. Publ. Serv., Los Alamitos, California, **2009**, pp. 1–7.

**Table 1.** The initial concentration at logic **1** (for the inputs) and values (for the filtering compound, when added) used in various experiments as identified in the text and figures.

| $[H_2O_2]_1$ | $[D]_1$ | | | $[F]$ | |
|---|---|---|---|---|---|
| 20.0 µM | [ABTS](0) = | | 250 µM | [Asc](0) = | 17.5 µM |
| 3.00 mM | [K$_4$Fe(CN)$_6$](0) = | | 30.0 mM | [Asc](0) = | 4.00 mM |
| 500 µM | [guaiacol](0) = | | 1.00 mM | [Asc](0) = | 250 µM or 120 µM |
| 500 µM | [guaiacol](0) = | | 1.00 mM | [GSH](0) = | 120 µM |
| 500 µM | [guaiacol](0) = | | 1.00 mM | [NADH](0) = | 120 µM |

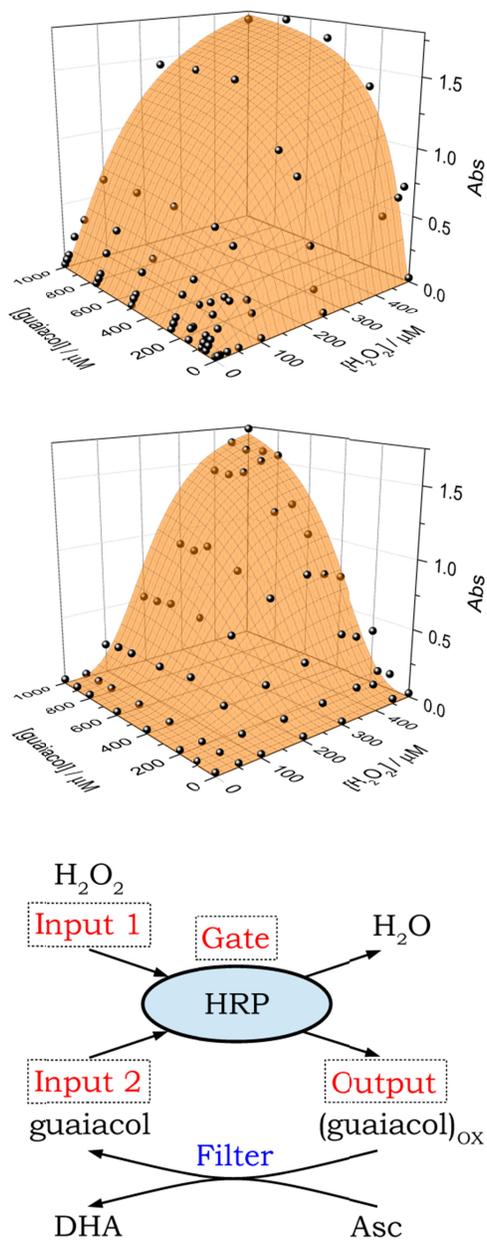

**Figure 1.** Illustration of the convex response and its conversion to double-sigmoid by filtering, for our favorable system, as described in Section 4, with guaiacol as the chromogen, and with [Asc](0) = 0 (top panel: no filtering) or 120 μM (middle panel: with filtering). Heavy dots show experimental data, whereas the surfaces represent model fitting. The other experimental parameters are described in the text. The scheme (bottom panel) sketches the biocatalytic process and the added filtering reaction (where DHA stands for dehydroascorbate, which is the product of Asc oxidation).

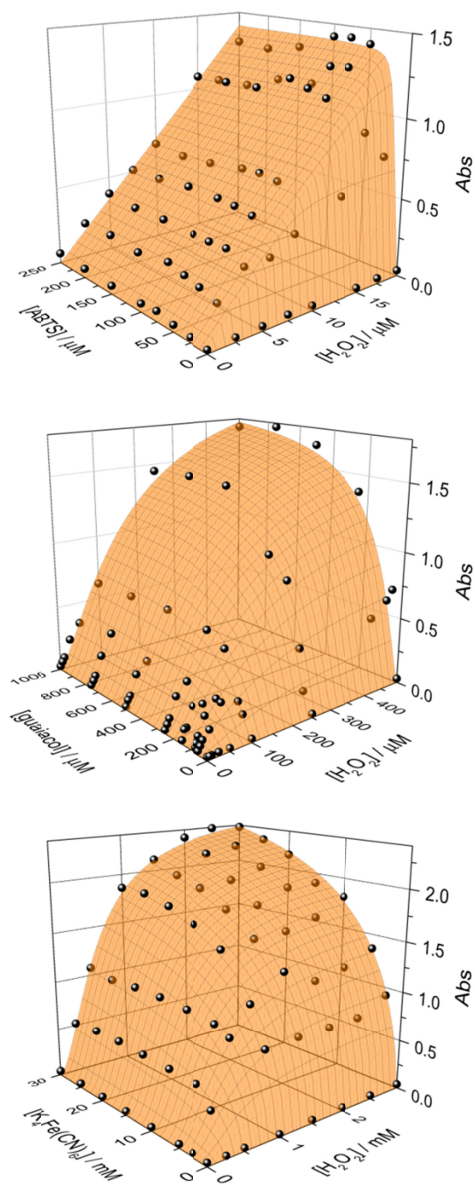

**Figure 2.** Experimental data for the absorbance (Abs), plotted as heavy dots, for gate time $t_g$ = 900 s, for the processes without filtering, with ABTS (top panel), guaiacol (middle panel, identical to the top panel in Figure 1), $K_4Fe(CN)_6$ (bottom panel) as chromogens. The surfaces represent the two-parameter fit according to the rate equations introduced in Section 3.

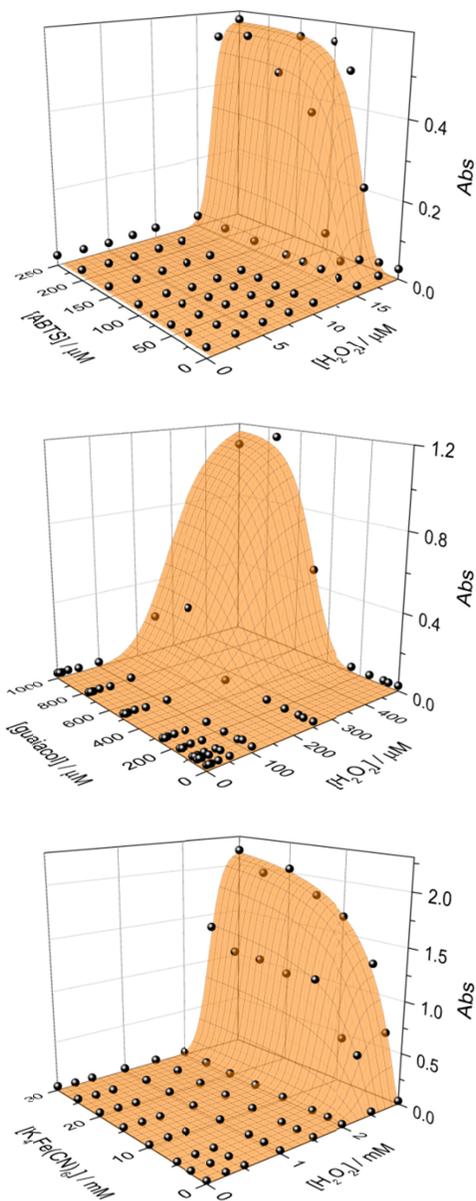

**Figure 3.** Experimental data for the absorbance (Abs), plotted as heavy dots, for gate time $t_g$ = 1800 s, for the processes with intense filtering, with ABTS (top panel), guaiacol (middle panel, with a more intense filtering than in the case shown in Figure 1), $K_4Fe(CN)_6$ (bottom panel) as chromogens, and with Asc as the filtering chemical, with, respectively, [Asc](0) = 17.5, 250, 4000 μM.

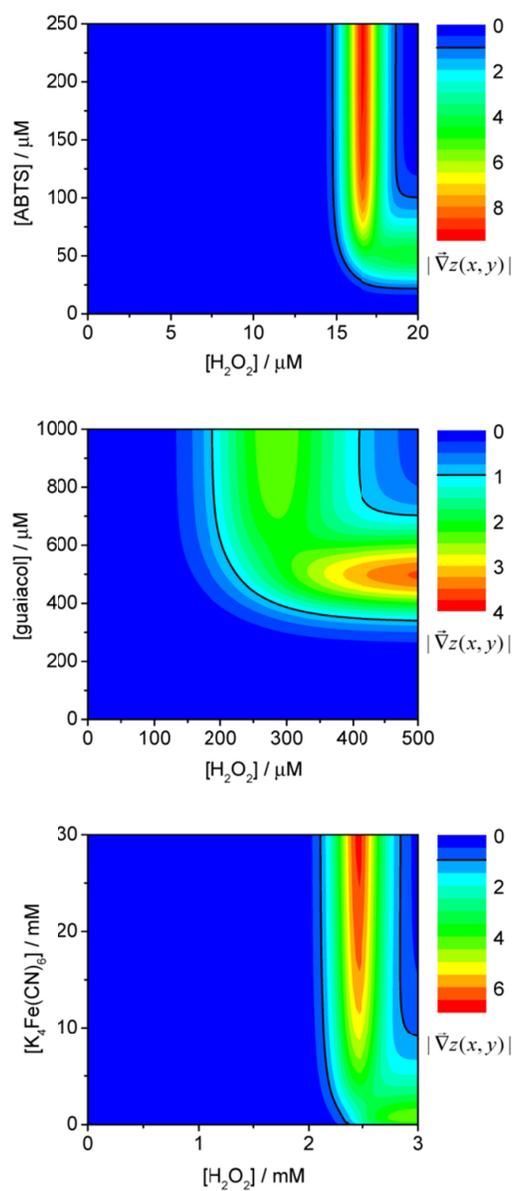

**Figure 4.** Color-coded contour plots of the absolute value of the gradient vector (the slope) in terms of the rescaled variables, $|\vec{\nabla}z(x,y)|$, calculated from the fitted model surfaces for the three experimentally realized "over-filtered" systems, ABTS, guaiacol, and $K_4Fe(CN)_6$ (top to bottom, in the same order as in Figure 3). The black solid lines mark the contours with slope $|\vec{\nabla}z(x,y)| = 1$. Note that due to the variations in the ranges of the slope values, the color codings of the panels are different.

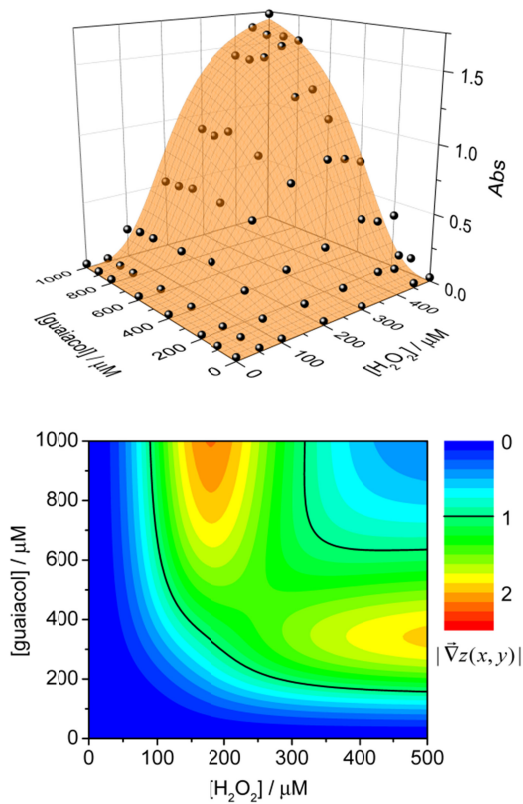

**Figure 5.** Experimental data for the absorbance for gate time $t_g$ = 1800 s, for the processes with guaiacol as the chromogen, and with [Asc](0) = 120 μM for filtering (top panel, identical to the middle panel in Figure 1). Also shown is a color-coded contour plot of the corresponding slope, calculated from the fitted model surface (bottom panel).

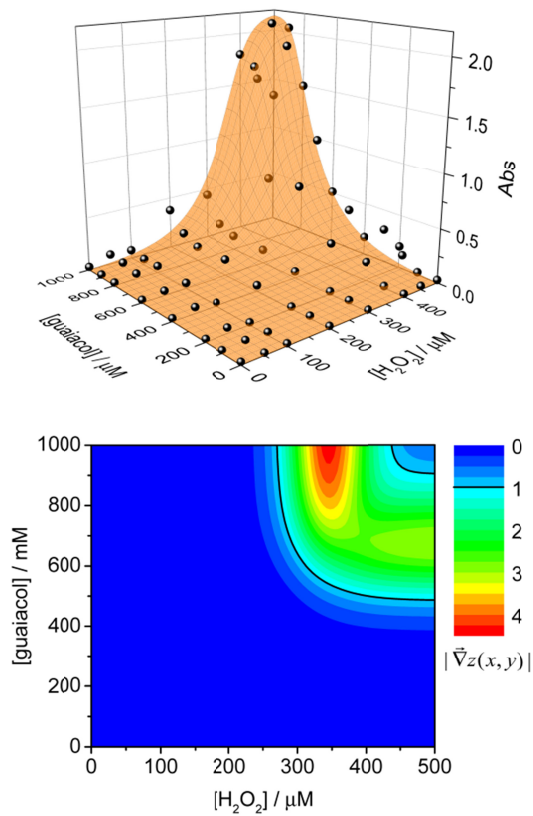

**Figure 6.** Experimental data for the absorbance for gate time $t_g$ = 1800 s, for the processes with guaiacol as the chromogen, and with [NADH](0) = 120 µM for filtering (top panel). Also shown is a color-coded contour plot of the corresponding slope, calculated from the fitted model surface (bottom panel).

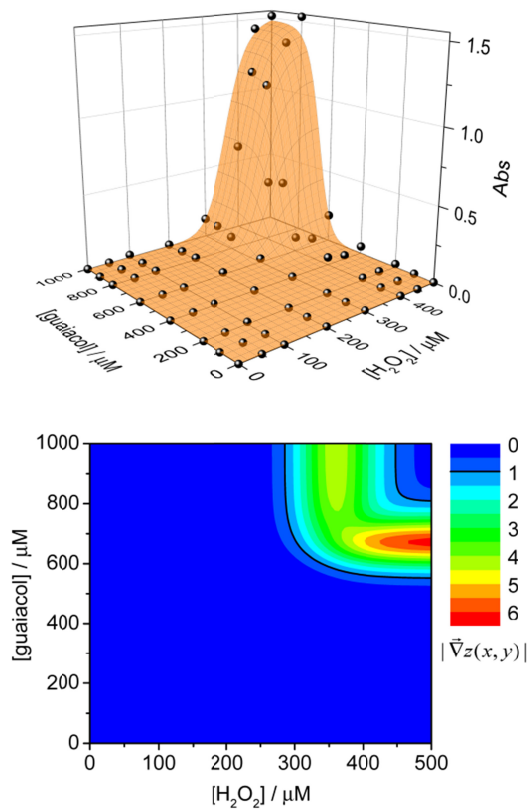

**Figure 7.** Experimental data for the absorbance for gate time $t_g$ = 1800 s, for the processes with guaiacol as the chromogen, and with [GSH](0) = 120 μM for filtering (top panel). Also shown is a color-coded contour plot of the corresponding slope, calculated from the fitted model surface (bottom panel).